\documentclass[12pt]{article}

\usepackage{amsmath,amssymb}
\usepackage{slashed}
\usepackage{xcolor} %Define colors
\usepackage{graphicx}
\usepackage{url}
\usepackage{cite}

\usepackage{verbatim}
   \allowdisplaybreaks
\makeatletter
\@addtoreset{equation}{section}
\makeatother

\graphicspath{{./Figures/}}
\newcommand{\be}{\begin{equation}} 
\newcommand{\ee}{\end{equation}} 
\newcommand{\bea}{\begin{eqnarray}}  
\newcommand{\eea}{\end{eqnarray}}
\newcommand{\bs}{\begin{split}} 
\newcommand{\es}{\end{split}}

\newcommand{\tr}{\operatorname{tr}}

\begin{document}
\thispagestyle{empty}

\begin{center}
%\hfill CERN-PH-TH-2015-124
%\hfill UAB-FT-xxx
\begin{center}

\vspace{.5cm}

{\Large\bf Strong electroweak phase transition from\\ \vspace{0.3cm}Supersymmetric Custodial Triplets}

\end{center}
\vspace{1.cm}

\textbf{Mateo Garcia-Pepin$^{\,a}$,
and Mariano Quiros$^{a,\, b}$}\\

\vspace{.1cm}
${}^a\!\!$ {\em {Institut de F\'{\i}sica d'Altes Energies (IFAE),\\ The Barcelona Institute of  Science and Technology (BIST),\\ Campus UAB, 08193 Bellaterra (Barcelona) Spain}}

\vspace{.1cm}
${}^b\!\!$ {\em {Instituci\'o Catalana de Recerca i Estudis  
Avan\c{c}ats (ICREA), \\ Campus UAB, 08193 Bellaterra (Barcelona) Spain}}

\end{center}

\vspace{0.8cm}

\centerline{\bf Abstract}
\vspace{2 mm}
\begin{quote}%\small
The Supersymmetric Custodial Triplet Model, a supersymmetric generalization of the Georgi-Machacek model, has proven to be an interesting modification of the MSSM. It extends the MSSM Higgs sector by three extra $SU(2)_L$ triplets in such a way that approximate custodial invariance is preserved and $\rho$-parameter deviations are kept under control. By means of a sizeable triplet contribution to electroweak breaking the model is able to generate a barrier at tree level between the false vacuum and the electroweak one. This will result in a strong first order phase transition for an important region of the parameter space. We also look at the gravitational waves that could be generated as a result of the phase transition and show how future interferometers could be used as a probe of the model.

\end{quote}

\vfill

\newpage

%%%%%%%%%%%%%%%%%%%%%%%%%%%%%%%%%%
%%%%%%%%%%%%%%%%%%%%%%%%%%%%%%%%%%
\section{Introduction} \label{sec:intro} 
Electroweak baryogenesis (EWBG) is an interesting mechanism that could explain the observed asymmetry between matter and antimatter in the universe~\cite{Sakharov:1967dj} (for reviews see~\cite{Cohen:1993nk,Quiros:1994dr,Quiros:1999jp,Rubakov:1996vz,Carena:1997ys,Morrissey:2012db}). It ties together cosmology and physics at the electroweak (EW) scale, specifically the process of electroweak symmetry breaking (EWSB). However for this scenario to work, the electroweak phase transition (EWPT) needs to be a strong first order one, i.e.~it should proceed through bubble nucleation and sphaleron transitions should be sufficiently suppressed in the broken phase. The latter point prevents the asymmetry generated in the bubble walls to not be washed out once the broken phase fills up the universe.

With a $125$ GeV Higgs the Standard Model (SM) potential does not feature a barrier between unbroken and broken phases at zero temperature, although this barrier could in principle be produced by temperature dependent contributions to the potential via cubic terms. However, the SM degrees of freedom are not sufficient to generate a large enough barrier~\cite{Anderson:1991zb}. One then needs to add light degrees of freedom beyond the SM ones to radiatively generate the barrier, since heavy degrees of freedom decouple from the thermal bath and only light states provide a non negligible contribution to the effective potential.

Moreover the Sakharov conditions for successful baryogenesis require a much larger amount of CP violation than the one present in the SM and one needs to find sources beyond (BSM). In principle, supersymmetric extensions of the SM, such as the MSSM, could provide the required amount. Then, if one wants to embed EWBG in a supersymmetric context one should first check whether or not the MSSM features a strong enough first order phase transition. In principle the MSSM is able to generate a first order EWPT by the introduction of light stops which can generate large cubic terms at finite temperature. The problem in this case is that stops are required to be really light (below $\sim 150$ GeV~\cite{Carena:2008vj}) and unless one goes to very special models this mass range for stops is excluded by experimental searches. Moreover such light degrees of freedom modify the Higgs couplings and we would have seen these modifications by now~\cite{Curtin:2012aa,Carena:2012np,Katz:2014bha}. The problem can be generalized to any BSM proposal that tries to generate a first order EWPT radiatively: new light degrees of freedom below experimental bounds are commonly required and it often becomes difficult to accommodate a strong enough first order EWPT with collider searches. 

An interesting approach that can be taken is to modify the tree level potential and try to generate a barrier already at $T=0$. This can be done by extending the Higgs sector of the MSSM, for instance e.g.~by adding a gauge singlet field, the NMSSM~\cite{Pietroni:1992in}. To avoid possible problems with tadpole generation and/or domain wall problems of the singlet field~\cite{Bagger:1995ay,Abel:1995wk}, here we use the same approach but using the Supersymmetric Custodial Triplet Model (SCTM) which extends the field content of the MSSM by three $SU(2)_L$ chiral superfields and was first introduced in~\cite{Cort:2013foa}. The model makes use of the custodial symmetry to solve the $\rho$-problem of theories with triplets, and it is able to raise the tree level Higgs mass through new $F$-term contributions and fit the $\sim 125$ GeV measurement without the need of super-heavy stops. At the same time it generates largish triplet vacuum expectation values (VEVs) that can participate in the EW breaking up to a $\sim 15\%$ order. This latter fact has a wide variety of theoretical and phenomenological consequences that have been studied in several publications~\cite{Garcia-Pepin:2014yfa,Delgado:2015bwa,Delgado:2015aha}. One of its most interesting features, as we will see in this paper, is that it is able to generate a barrier between the origin and the EW minimum already at tree level. In this paper we explore this fact and analyze the behavior of its EWPT for the purpose of being able to generate a successful EWBG in supersymmetric extensions of the SM.   

The paper is organized as follows: In Section~\ref{sec:model} we introduce the model and the loop corrections to the potential, both at zero temperature and the temperature dependent ones. In Section~\ref{sec:strength} we study the strength of the phase transition at the degeneracy temperature. Section~\ref{subsec:tunneling} is devoted to the study of the thermal tunneling and the nucleation temperature. In Section~\ref{sec:pheno} we study the gravitational waves generated by the EWPT of the model. We end with a summary of our work and conclusions.

%%%%%%%%%%%%%%%%%%%%%%%%%%%%%%%%%%
%%%%%%%%%%%%%%%%%%%%%%%%%%%%%%%%%%
\section{The Model} 
\label{sec:model}
In this section we will construct a supersymmetric Higgs sector which is manifestly invariant under the global symmetry $SU(2)_L\otimes SU(2)_R$. The MSSM Higgs sector $H_1$ and $H_2$ with respective hypercharges $Y=(-1/2,\,1/2)$ 
 \be
   H_1=\left( \begin{array}{c}H_1^0\\ H_1^-\end{array}\right),\quad
   H_2=\left( \begin{array}{c}H_2^+\\ H_2^0\end{array}\right)
   \ee
 is complemented with $SU(2)_L$ triplets, $\Sigma_{Y}$, with hypercharges  $Y=(-1,\, 0,\, 1)$ 
 \be
 \Sigma_{-1}=\left(\begin{array}{cc} \frac{\chi^-}{\sqrt{2}} & \chi^0\\\chi^{--}& -\frac{\chi^-}{\sqrt{2}}
 \end{array}
 \right),\quad  \Sigma_{0}=\left(\begin{array}{cc} \frac{\phi^0}{\sqrt{2}} & \phi^+\\ \phi^{-}& -\frac{\phi^0}{\sqrt{2}}
 \end{array}
 \right),\quad  \Sigma_{1}=\left(\begin{array}{cc} \frac{\psi^+}{\sqrt{2}} & \psi^{++}\\\psi^{0}& -\frac{\psi^+}{\sqrt{2}}
 \end{array}
 \right)\ ,
 \ee
where $Q=T_{3L}+Y$.

The two doublets and the three triplets are organized under $SU(2)_L\otimes SU(2)_R$ as $\bar H=(\textbf{2},\bar {\textbf{2}})$, and $\bar \Delta=(\textbf{3},\bar{\textbf{3}})$ where
\be
\bar H=\left( \begin{array}{c}H_1\\ H_2\end{array}\right),\quad
\bar\Delta=\left(\begin{array}{cc} -\frac{\Sigma_0}{\sqrt{2}} & -\Sigma_{-1}\\ -\Sigma_{1}& \frac{\Sigma_0}{\sqrt{2}}\end{array}\right)
\ee
and $T_{3R}=Y$.   The invariant products for doublets $A\cdot B\equiv A^a\epsilon_{ab}B^b$  and anti-doublets $\bar A\cdot \bar B\equiv\bar A_a\epsilon^{ab}\bar B_c$ are defined by $\epsilon_{21}=\epsilon^{12}=1$. \\
The $SU(2)_L\otimes SU(2)_R$ invariant superpotential is defined as
\be
W_0=\lambda \bar H\cdot \bar\Delta\bar H+\frac{\lambda_3}{3}\tr\bar\Delta^3+\frac{\mu}{2}\bar H\cdot\bar H+\frac{\mu_\Delta}{2}\tr \bar\Delta^2
\label{W0}
\ee
\subsection{Scalar potential at zero temperature and the vacuum}
\label{section:minim}

Accordingly with the previous superpotential and gauge particle content the total tree level potential, as dictated by the symmmetries of the theory, is given by
\be
V_{\rm{tree}}=V_F+V_D+V_{\rm soft} \, ,
\ee
where
\begin{eqnarray}
V_{\rm soft}&=&m_{H_1}^2|H_1|^2+m_{H_2}^2|H_2|^2+m_{\Sigma_1}^2 \tr |\Sigma_1|^2+m_{\Sigma_{-1}}^2 \tr |\Sigma_{-1}|^2+m_{\Sigma_0}^2 \tr |\Sigma_0|^2
\nonumber\\
&+&\left\{\frac{1}{2}m_3^2\bar H\cdot\bar H
+ \frac{1}{2}B_\Delta\tr\bar\Delta^2+A_\lambda \bar H\cdot \bar\Delta \bar H+\frac{1}{3}A_{\lambda_3}\tr\bar\Delta^3+h.c.\right\} \, .
\label{Vsoft}
\end{eqnarray}
Note that the soft part of the potential we just wrote is the same as in~\cite{Cort:2013foa} but with non custodial soft masses that explicitly spoil the $SU(2)_L\otimes SU(2)_R$ invariance. This small breaking of custodial invariance can be understood as coming from the running of the model parameters from the scale $M$, where supersymmetry is broken and the theory is defined as exactly custodial in the Higgs sector, to the weak scale where the model parameters are defined~\cite{Garcia-Pepin:2014yfa}. This small breaking of custodial invariance is accounted for in the minimization process next described.

To the tree level piece one has to add the Coleman-Weinberg contribution for the one-loop radiative corrections at $T=0$, which will depend on the considered background scalar fields: $H_1^0,H_2^0$ from the usual MSSM $SU(2)_L$ doublets, and $\psi^0,\phi^0,\chi^0$, corresponding to the new triplet sector. We will work for simplicity in the $\overline{MS}$ renormalization scheme for which
\begin{equation}
\Delta V_1^{T=0}(\phi_k) = \sum_i \frac{n_i}{64\pi^2}m_i^4(\phi_k)\left( \log{\frac{m_i^2(\phi_k)}{Q^2}}-C_i\right)\, ,
\end{equation}
where $C_i=5/6$ for gauge bosons and $C_i=3/2$ for the rest of states and $n_i$ is the number of degrees of freedom for each particle ($n_W=6$, $n_Z=3$, $n_t=-12$, $n_{\tilde{t}_1}=6$, $n_{\tilde{t}_2}=6$, \dots) and we write $\phi_k\equiv H_1^0,H_2^0,\psi^0,\phi^0,\chi^0$ for simplicity. In the $\overline{MS}$ (as in any mass independent renormalization scheme) decoupling of heavy particles is not automatically implemented, but has to be done by hand at a scale of the order of their mass where they are integrated out, eventually leaving some threshold corrections (the run-and-match procedure) in the low energy effective theory. The run-and-match procedure guarantees the absence of large logarithms in the effective potential (for useful examples of this procedure in the MSSM see Refs.~\cite{Carena:2008rt,Masina:2015ixa}). On the other hand the $\overline{MS}$ renormalization scheme changes the location of the tree-level potential minimum as well as the value of the (running) Higgses masses. In other words the tree-level potential must be minimized after inclusion of radiative corrections, as we will do next.

The total background-dependent one-loop zero temperature potential is then
\begin{equation}
V_1(\phi_k)=V_{\rm{tree}}(\phi_k) + \Delta V_1^{T=0}(\phi_k)
\end{equation}
and the EWSB vacuum is derived by solving the five minimization conditions
\begin{equation}
\left.\frac{\partial V_1(\phi_k)}{\partial H_1^0}\right|_{\phi_k=v_k}
= \left.\frac{\partial V_1(\phi_k)}{\partial H_2^0}\right|_{\phi_k=v_k}
=\left. \frac{\partial V_1(\phi_k)}{\partial \psi^0}\right|_{\phi_k=v_k}
= \left.\frac{\partial V_1(\phi_k)}{\partial \phi^0}\right|_{\phi_k=v_k}
= \left.\frac{\partial V_1(\phi_k)}{\partial \chi^0}\right|_{\phi_k=v_k}=0\, ,
\end{equation}
where we impose the EW vacuum to be at
\be
\label{vacuum}
 v_1=\sqrt{2}\cos{\beta}\,v_H,\quad v_2=\sqrt{2}\sin{\beta}\,v_H\quad \textrm{and}\quad v_\psi=v_\chi=v_\phi\equiv v_\Delta.
\ee 
so that we allow breaking of custodial invariance only in the doublet sector, which is a very good approximation as that breaking is triggered in the running mainly by the top Yukawa coupling~\cite{Garcia-Pepin:2014yfa}. The Higgs mass is computed numerically from the scalar mass matrix that is derived from the above potential, and we have checked that it is very well approximated by the analytical expressions from Refs.~\cite{Carena:1995bx,Carena:1995wu}, although the plots are based on the numerical calculation. Note that we are only including dominant contributions to the Higgs mass~\footnote{Because we have introduced three extra $SU(2)_L$ triplets, the scalar sector of this model is enhanced with respect to the MSSM by a new set of states that carry a large triplet component and couple very weakly~\cite{Delgado:2015bwa}.}. 

The custodial symmetry of the vacuum is only broken by $\tan\beta$. As it was argued in~\cite{Delgado:2015aha}, by allowing for $\tan{\beta}\neq 1$ to deal with the parametrization of some possible custodial breaking, we capture the main features of it without the need to perform a thorough study of a UV complete model. To set the $Z$ mass, the total VEV must be
\be
\label{vev174}
v^2 \equiv (174~\mathrm{GeV})^2= 2v_H^2+8v_{\Delta}^2.
\ee
Finally, in order to solve the five minimization conditions we need to fix five parameters. We will choose for them the soft scalar masses $m_{H_1},m_{H_2}$ and $m_{\Sigma_1},m_{\Sigma_{-1}},m_{\Sigma_0}$ as in Ref.~\cite{Delgado:2015aha}. 

\subsection{Finite temperature scalar potential} \label{sec:fintemp} 

The finite temperature potential at one-loop is
\begin{equation}
V_1(\phi_k,T)=V_{\rm{tree}}(\phi_k) + \Delta V_1^{T=0}(\phi_k) + \Delta V_1(\phi_k,T)+\Delta V_{\mathrm{daisy}}(\phi_k,T)
\end{equation}
with the finite temperature part
\begin{equation}
\label{finitetemp}
\Delta V_1(\phi_k,T)=\frac{T^4}{2\pi^2}\left( \sum_i n_i J_i\left[\frac{m_i^2(\phi_k)}{T^2}\right] \right)\, .
\end{equation}
and the thermal integrals~\footnote{These integrals can also be written in terms of an infinite sum of Bessel functions~\cite{Anderson:1991zb}
$$
J_{\pm}(y)\equiv -\sum_{n=1}^{\infty}\frac{(\pm 1)^n}{n^2}\,y^2 K_2\left(n y\right)\ .
$$
By truncating the sum to a large enough order, one can obtain a more calculable situation which still represents a good approximation to the thermal integrals written above. We will not use any high (low) temperature expansion in this work since our interesting parameter space does not qualify for any of the two regimes.},
\be
J_{\pm}(y)\equiv \int_0^{\infty} dx \,x^2 \log{\left(1\mp e^{-\sqrt{x^2+y}}\right)}
\ee 
Here $J_i=J_+(J_-)$ if the $i^{th}$ particle is a boson (fermion). The Daisy piece is given by
\be
\Delta V_{\mathrm{daisy}}(\phi_k,T)=-\frac{T}{12\pi}\sum_{i=\mathrm{bosons}} n_i\left[\mathcal{M}_i^3(\phi_k,T)-m(\phi_k)^3 \right] \, ,
\ee
where
\be
\mathcal{M}_i^2=m_i^2(\phi_k)+\Pi_i(\phi_k,T)\, .
\ee
Since the thermal corrections to the (un-resummed) one-loop potential potential automatically decouple heavy degrees of freedom we will only Daisy resum the longitudinal components of  light gauge bosons $W_L$, $Z_L$ and $\gamma_L$ just as in the SM~\cite{Quiros:1999jp}. In the one-loop approximation 
\be
\begin{aligned}
\Pi_{W_T}(\phi_k,T)&=\Pi_{Z_T}(\phi_k,T)=\Pi_{\gamma_T}(\phi_k,T)=0\, , \\
\Pi_{W_L}(\phi_k,T)&=\frac{11}{6}g^2T^2 \ 
\end{aligned}
\ee
and the SM Debye masses $\mathcal{M}_i^2$ for $Z_L, \gamma_L$ are given by
\be
\begin{aligned}
\mathcal{M}_{Z_L}^2&=\frac{1}{2}\left( m_Z^2(\phi_k)+\frac{11}{6}\frac{g^2}{\cos^2{\theta_W}}T^2+\Delta(\phi_k,T) \right)\, , \\
\mathcal{M}_{\gamma_L}^2&=\frac{1}{2}\left( m_Z^2(\phi_k)+\frac{11}{6}\frac{g^2}{\cos^2{\theta_W}}T^2-\Delta(\phi_k,T) \right) \, .
\end{aligned}
\ee
Where
\be
\Delta^2(\phi_k,T)=m_Z^4(\phi_k)+\frac{11}{3}\frac{g^2\cos^2{2\theta_W}}{\cos^2{\theta_W}}\left( m_Z^2(\phi_k) +\frac{11}{12}\frac{g^2}{\cos^2{\theta_W}}T^2 \right)T^2\ .
\ee

%%%%%%%%%%%%%%%%%%%%%%%%%%%%%%%%%%
%%%%%%%%%%%%%%%%%%%%%%%%%%%%%%%%%%

\section{Strength of the phase transition} \label{sec:strength} 

We have found that $\mu$ and $\mu_\Delta$ are the parameters to which the potential shows more sensitivity for creating a barrier between the origin and the EW minimum already at $T=0$, they are therefore critical to the study of the phase transition. To simplify the study we will make contour plots of different quantities on the $(\mu,\mu_\Delta)$ plane while holding other parameters fixed. To start doing numerical computations we first choose a set of benchmark values given by
\be
\begin{aligned}
A_{\lambda}&=A_{\lambda_3}=A_t=0, \quad \lambda_3=0.35, \\ m_3 &= 750~\mathrm{GeV}, \quad
B_{\Delta}=-(750 ~\mathrm{GeV})^2,\\
m_{\tilde{Q}_3}&= 800~\mathrm{GeV}, ~~\text{and}~~ m_{\tilde{u}_3^c} = 800~\mathrm{GeV}.
\label{eqn:benchmark}
\end{aligned}
\ee

\begin{figure}[htb]
\begin{center}
\includegraphics[scale=.91]{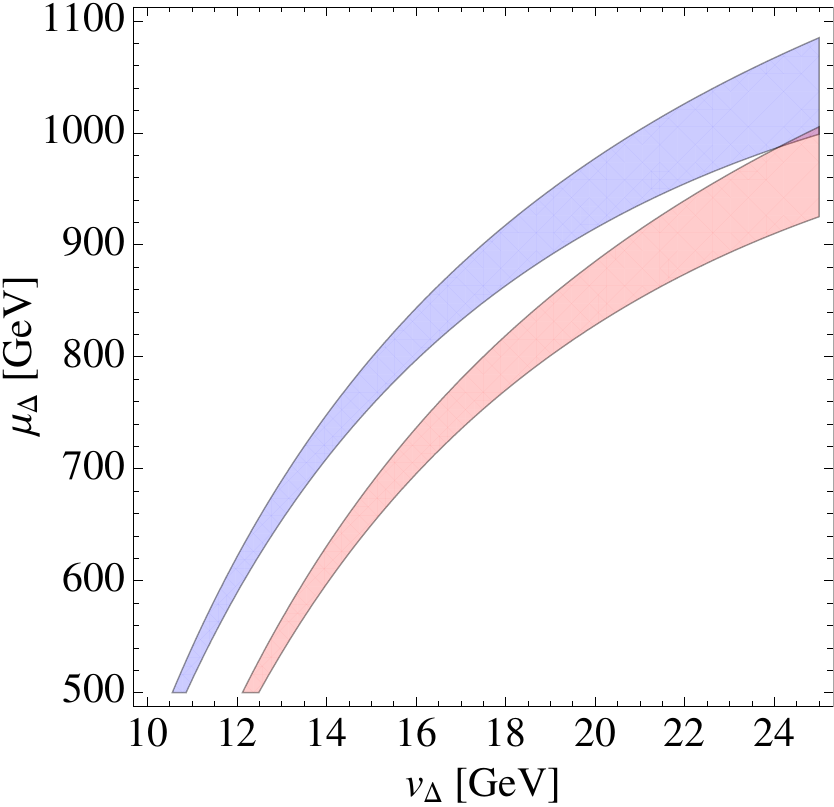}
\includegraphics[scale=.52]{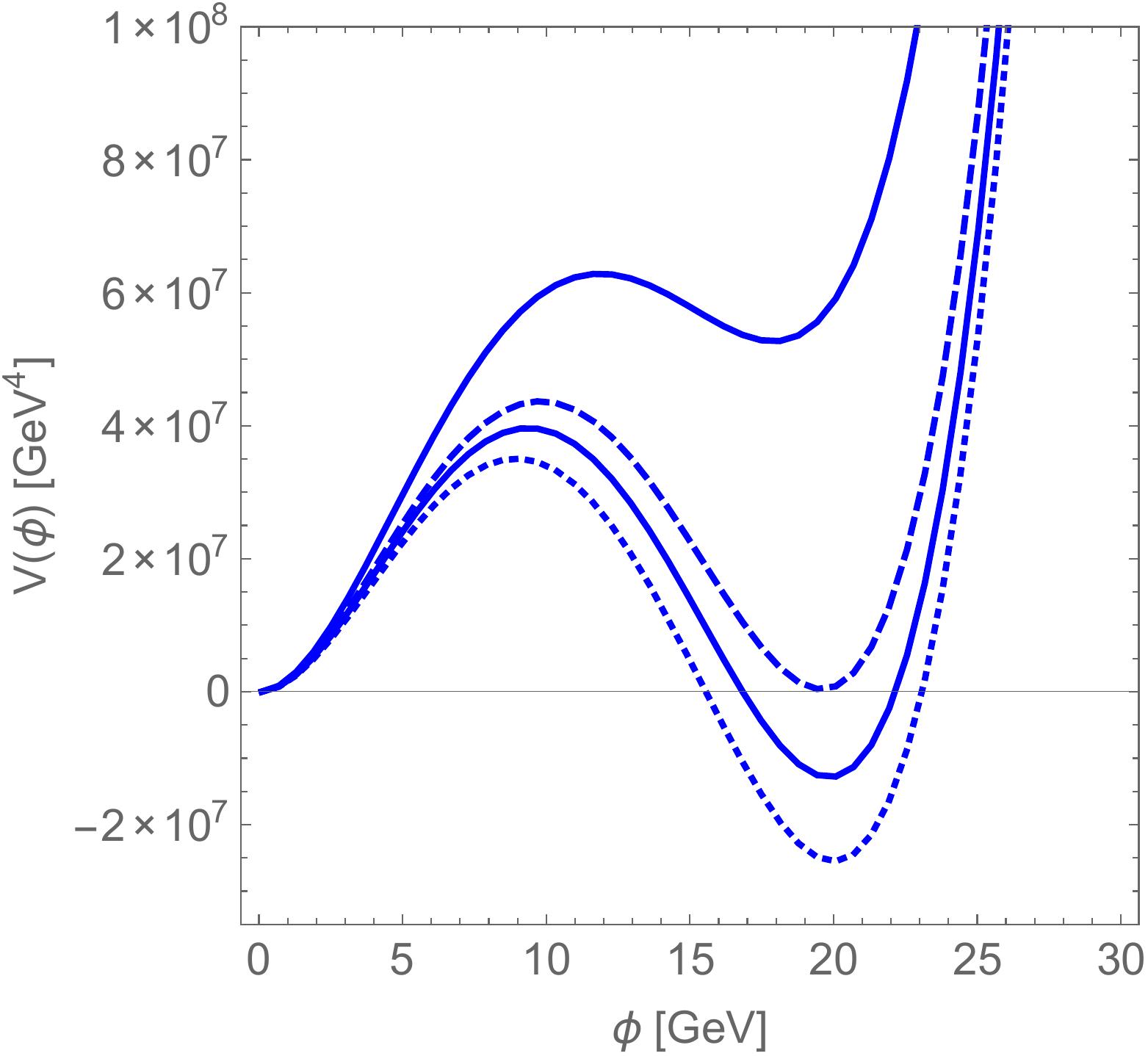}
\end{center}
\caption{~\emph{Left panel: For $\lambda=0.7$, $\mu=750$ GeV, and $\tan{\beta}=1$ (blue), $\tan{\beta}=1.5$ (red), regions in the $(v_\Delta, \mu_\Delta)$ plane where the zero temperature tree level potential shows a false minimum at the origin. Right panel: For $\tan{\beta}=1$, $v_\Delta=20$, GeV $\mu=650$ GeV and $\mu_\Delta=415$ GeV sections of the five-dimensional potential at different temperatures along the direction that joins the false and true vacuums by a straight line, $T=0$ and $T=T_c$ are depicted with dotted and dashed lines respectively.}}
\label{fig:spectrum}
\end{figure}

In the left panel of Fig.~\ref{fig:spectrum} we plot regions in the $(v_\Delta,\mu_\Delta)$ plane, for $\mu=750$ GeV and different values of $\tan\beta$, where the origin is a false minimum at zero temperature and therefore there is a barrier separating the origin from the true EW minimum. These regions are then eligible to generate, at finite temperature, a strong enough EWPT as that exhibited in the right panel of Fig.~\ref{fig:spectrum}. One can realize from the plot in the left panel of Fig.~\ref{fig:spectrum} that this region only appears, and becomes important, when $v_\Delta$ is non negligible. By means of the needed sizeable values of $v_\Delta$, the plot shows how critical is for the appearance of the barrier to have a non negligible contribution of the triplet sector to EWSB. 

\begin{figure}[htb]
\begin{center}
\includegraphics[scale=.47]{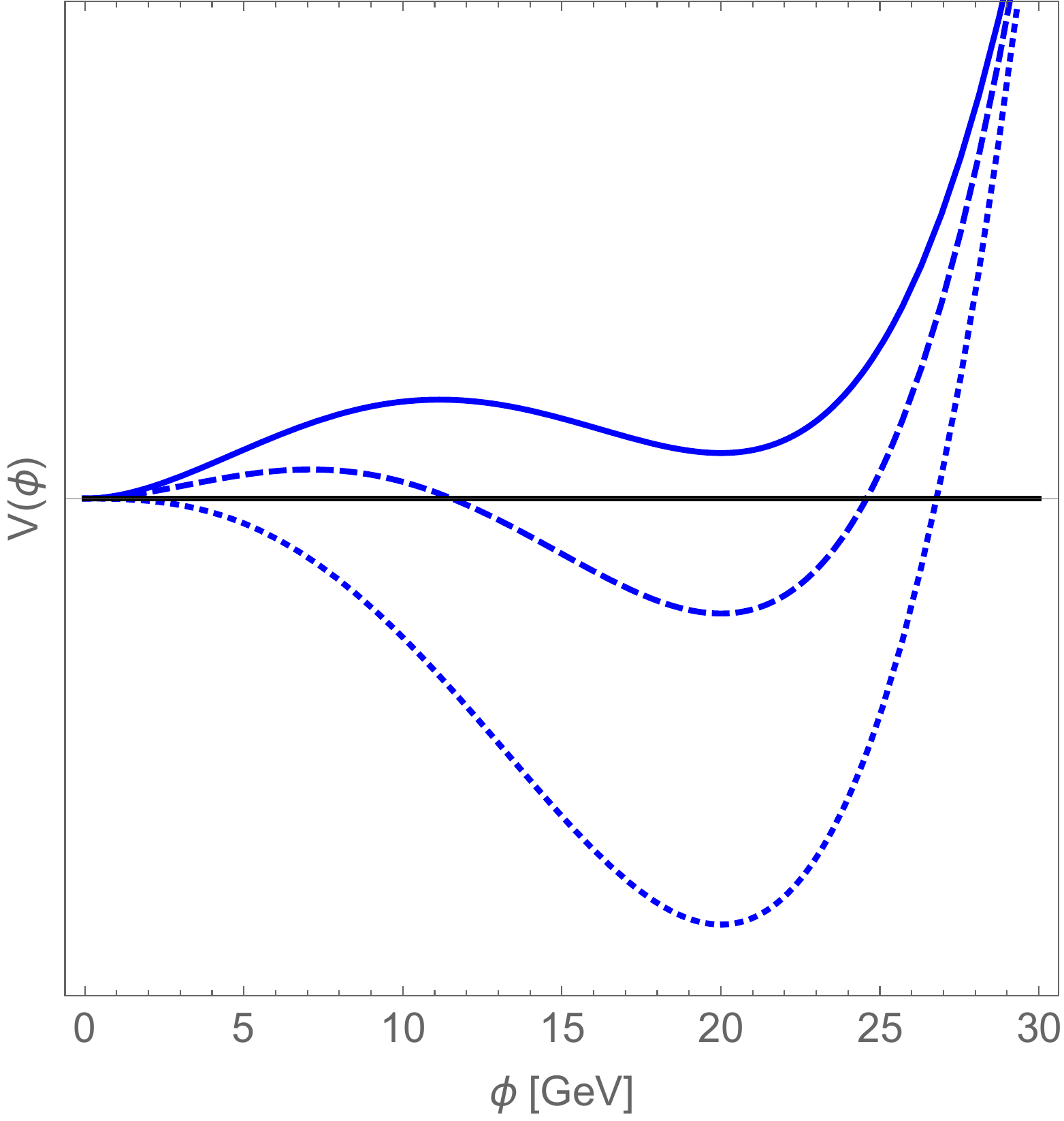}
\includegraphics[scale=.64]{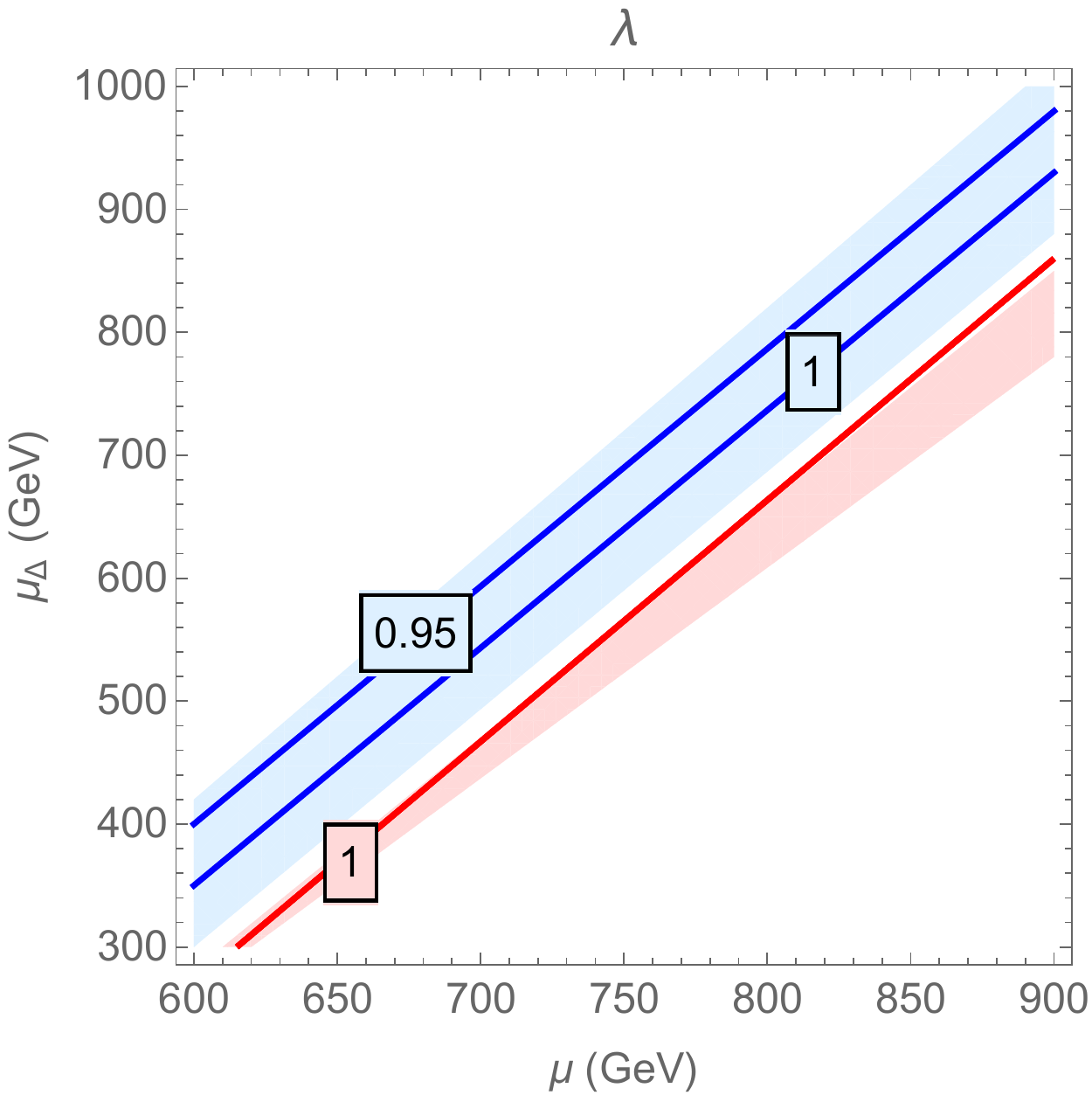}
\end{center}
\caption{~\emph{Left panel: Sections of the potential at zero temperature in the direction of minimum slope for values of $\mu$ and $\mu_\Delta$ above (dotted), inside (dashed) and below (solid) the bands where a first order phase transition is realized. Right panel: For $v_\Delta=20$ GeV, values of $\lambda$ that are needed to get the correct Higgs mass in the $(\mu,\mu_\Delta)$ plane for $\tan{\beta}=1$ (blue) and $\tan{\beta}=1.5$ (red).}}
\label{fig:lambda}
\end{figure}
For any fixed value of $\tan\beta$ and points outside the corresponding band the zero temperature potential does not fulfil our required conditions for having a strong enough EWPT at finite temperature. In particular for points below the corresponding band the EW vacuum is a false minimum (or even it does not exist) and thus no transition from the origin to the EW minimum is possible at any temperature. This is exhibited at a particular point below the band, for the zero temperature potential along the direction where the slope of the barrier is minimized, in the left panel of Fig.~\ref{fig:lambda} (solid line) where we can see that the EW minimum is not the true minimum. For points inside the corresponding band the EW minimum is the true minimum and the EWPT can proceed through a strong enough first order phase transition. The zero temperature potential for a point inside the band is exhibited in the left panel of Fig.~\ref{fig:lambda} (dashed line).
 Finally for points above the corresponding band, the origin of the zero temperature potential becomes a saddle point as shown in the left panel of Fig.~\ref{fig:lambda} (dotted line).
Therefore in this region the barrier between the origin and the EWSB minimum can only be generated by thermal corrections, and the EWPT is too weak (or not even first order) as it happens in the SM or in the MSSM.
At each point the value of the parameter $\lambda$ is adjusted such that the value of the Higgs mass reproduces the experimental result $m_H=125$ GeV. The needed values of $\lambda$ are provided in the right panel of Fig.~\ref{fig:lambda} where we show, for $v_\Delta=20$ GeV, in the $(\mu,\mu_\Delta)$ plane contour lines of constant values of $\lambda$ inside the bands  for $\tan{\beta}=1$ (blue) and $\tan{\beta}=1.5$ (red).

Once identified the region in the parameter space where our potential is able to generate a first order EWPT we will study its temperature dependence. We will search for points where the phase transition is strong enough as to avoid any washout of the generated baryon asymmetry due to sphaleron transitions. This condition translates into the following bound for the Standard Model~\cite{Farrar:1993hn},
\be 
\label{orderparam}
\frac{v(T_n)}{T_n} \gtrsim 1
\ee
where $v(T_n)$ is the VEV of the Higgs field at the nucleation temperature. We do not expect this bound to be very different in the present model, since the sphaleron energy is dominated by the contributions from the gauge field configurations excited in the sphaleron rather than the scalar ones~\cite{Klinkhamer:1984di}.

\begin{figure}[htb]
\begin{center}
\includegraphics[scale=.6]{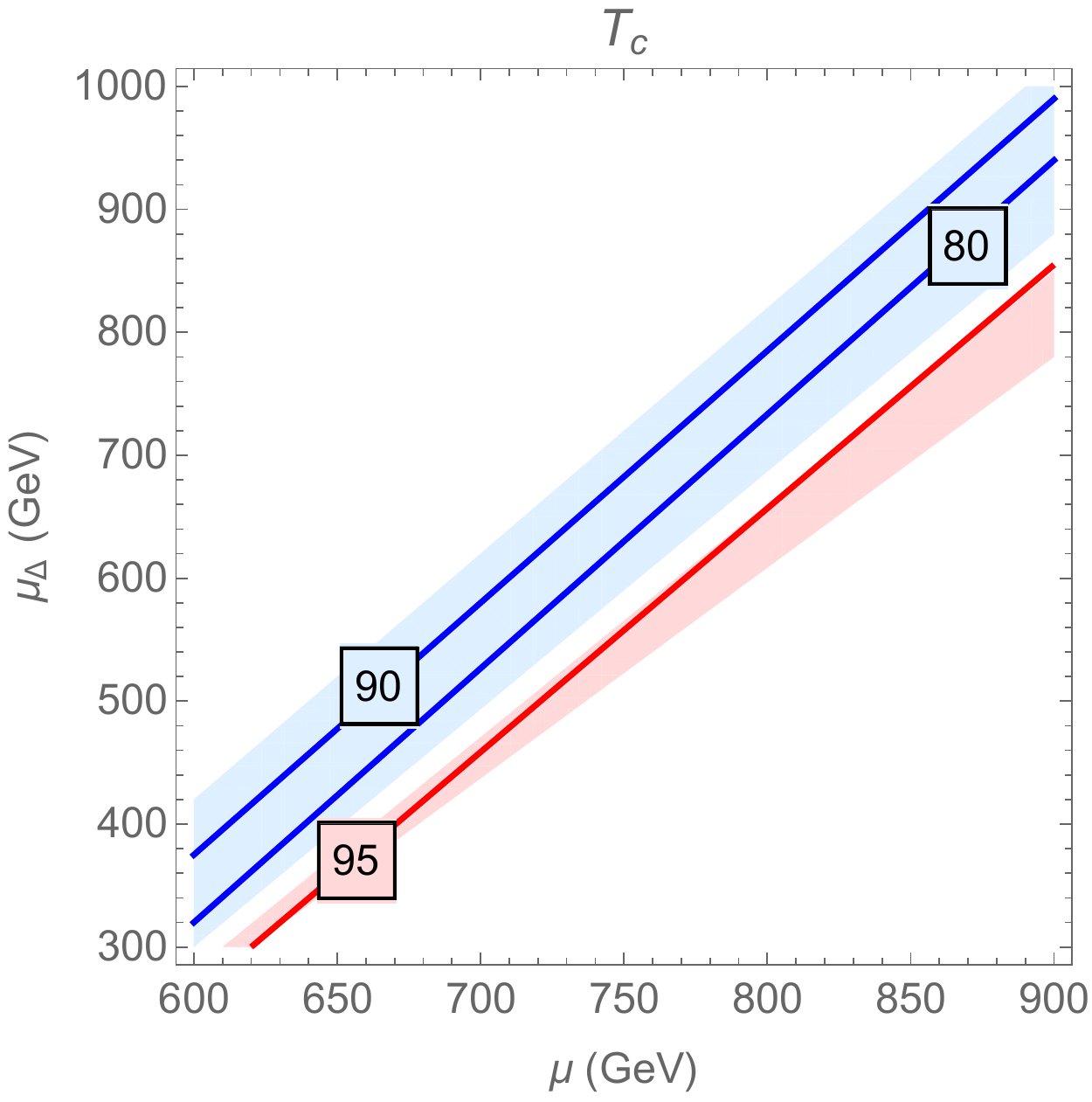}
\includegraphics[scale=.6]{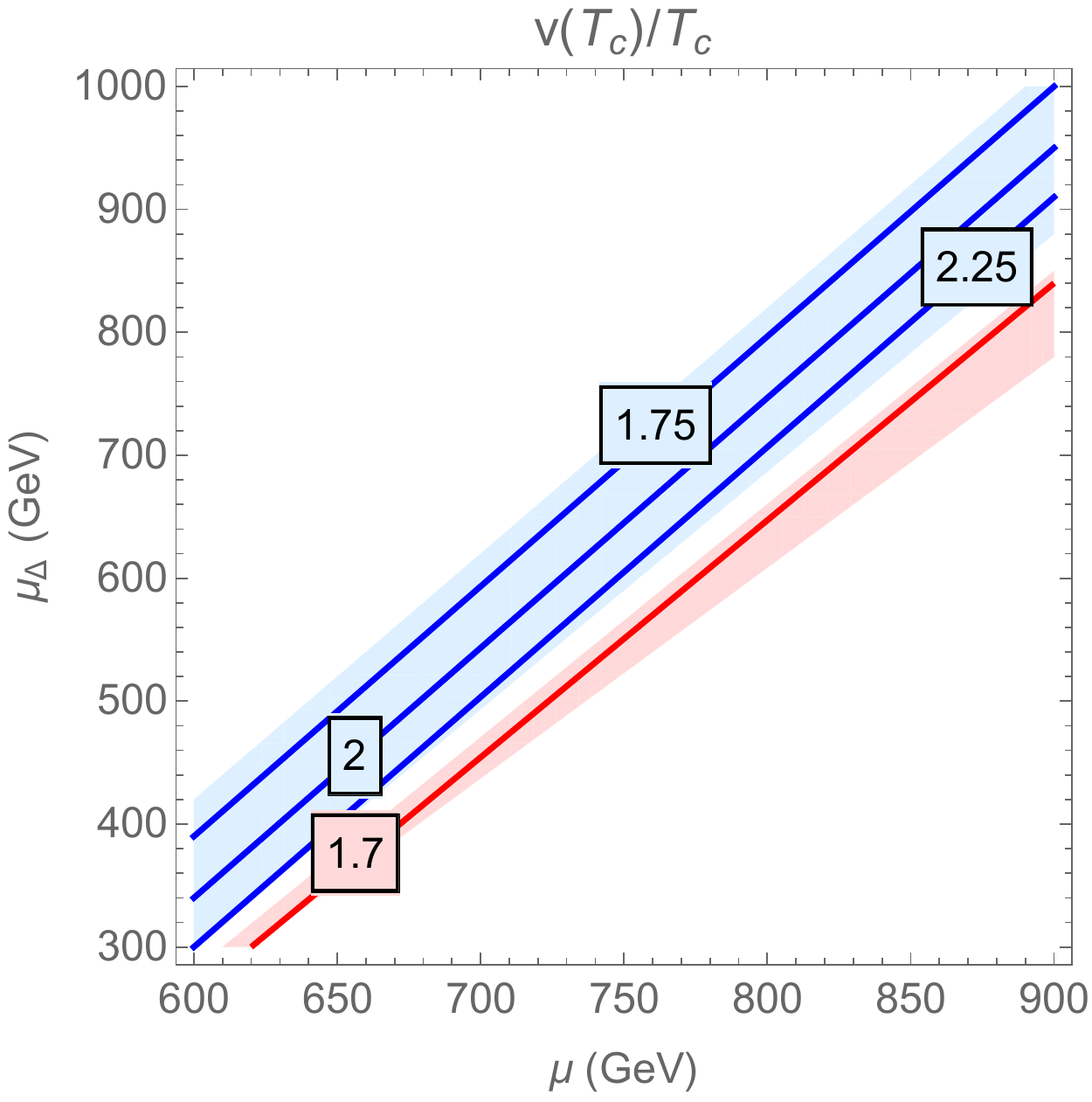}
\end{center}
\caption{~\emph{Critical temperature in GeV (left panel) and order parameter of the phase transition (right panel) in the $(\mu,\mu_\Delta)$ plane for $v_\Delta=20$ GeV and $\tan{\beta}=1$ (blue), $\tan{\beta}=1.5$ (red). In both plots we have shown the band where there is a barrier between the origin and the global EW minimum in the zero temperature effective potential for $\tan{\beta}=1$ (blue) and $\tan{\beta}=1.5$ (red). At each point $\lambda$ is adjusted such that the Higgs mass reproduces the experimentally observed value.}}
\label{fig:firstorder}
\end{figure}

As the condition $v(T_c)/T_c\lesssim v(T_n)/T_n$, where $v(T_c)$ defined by
\be 
v(T_c) = \sqrt{H_1^{0}(T_c)^2+H_2^{0}(T_c)^2+2\psi^0(T_c)^2 + 4\phi^0(T_c)^2 + 2\chi^0(T_c)^2}\ .
\ee
is the Higgs VEV at the critical temperature (the temperature at which both minima are degenerate), is generically satisfied as we will see later on in this paper, it is sufficient to consider the EWPT strong enough when the condition $v(T_c) \gtrsim T_c$ is fulfilled. In fact this sufficient condition is much simpler to analyze than (\ref{orderparam}) as it can (and will) be easily done in the full five-dimensional Higgs potential.

In Fig.~\ref{fig:firstorder} we present results for the critical temperature (left panel) and the order parameter of the phase transition at the critical temperature (right panel) in the $(\mu,\mu_\Delta)$ plane. Our results for the EWPT are even stronger than what it is shown in the left and right panels of Fig.~\ref{fig:firstorder}, since the true order parameter of the EWPT (the order parameter at the nucleation temperature) will be bigger than the one at the critical temperature, as it was already observed. We only show points where the strong phase transition is generated by the zero temperature potential exhibiting a false minimum at the origin, the blue (for $\tan\beta=1$) and red (for $\tan\beta=1.5$) bands. As we can see in the right panel of Fig.~\ref{fig:firstorder} the strength of the phase transition increases as we approach the lower boundary of the corresponding band. As we will see in section~\ref{sec:pheno} this region will be favored for the detection of the gravitational waves emitted during the EWPT.

%%%%%%%%%%%%%%%%%%%%%%%%%%%%%%%%%%%%%%%%
%%%%%%%%%%%%%%%%%%%%%%%%%%%%%%%%%%%%%%%%%
\section{Thermal tunnelling and nucleation temperature} \label{subsec:tunneling} 

Once we have computed the strength of the phase transition at the critical temperature, the next step is to compute the tunneling temperature to make sure that bubble nucleation does happen. Of course this is ensured if the phase transition is generated radiatively since there is no barrier at zero temperature and, as the universe cools down, we will always cross a point where the tunneling probability is $\mathcal O(1)$. However in the region we are interested in this is not guaranteed as there is a barrier at zero temperature and it could be too strong for the field to tunnel from the symmetric to the broken phase at any temperature. 

As the computation of the thermal tunneling in the five-field case presents computational challenges that are out of the scope of this work, we will use an approximation to strip down our five field configuration to a one-dimensional field space. We will first consider the following,
\be
H_1^0 \rightarrow \frac{v_1(T)}{v_2(T)}H_2^0\quad \textrm{and} \quad \psi^0 \rightarrow  \frac{v_\psi(T)}{v_\phi(T)}\phi^0,\, \chi^0 \rightarrow \frac{v_\chi(T)}{v_\phi(T)} \phi^0.
\label{aprox1}
\ee
For the doublet sector this approximation is expected to be a very good one near the decoupling limit, where all scalar masses are much heavier than the SM Higgs mass, which is nearby the spectrum we are considering in this paper~\footnote{For a light spectrum our calculation of the approximated nucleation temperature %$T_n^{app}$ 
might require strong corrections.}, for the dependence of $\tan\beta$ on the temperature is a mild one~\cite{Moreno:1998bq}. Moreover the smallness of $v_\Delta$ with respect to $v_H$ will also ensure that the triplet sector is well approximated by Eq.~(\ref{aprox1}). 

\begin{figure}[tbh]
\begin{center}
\includegraphics[scale=.6]{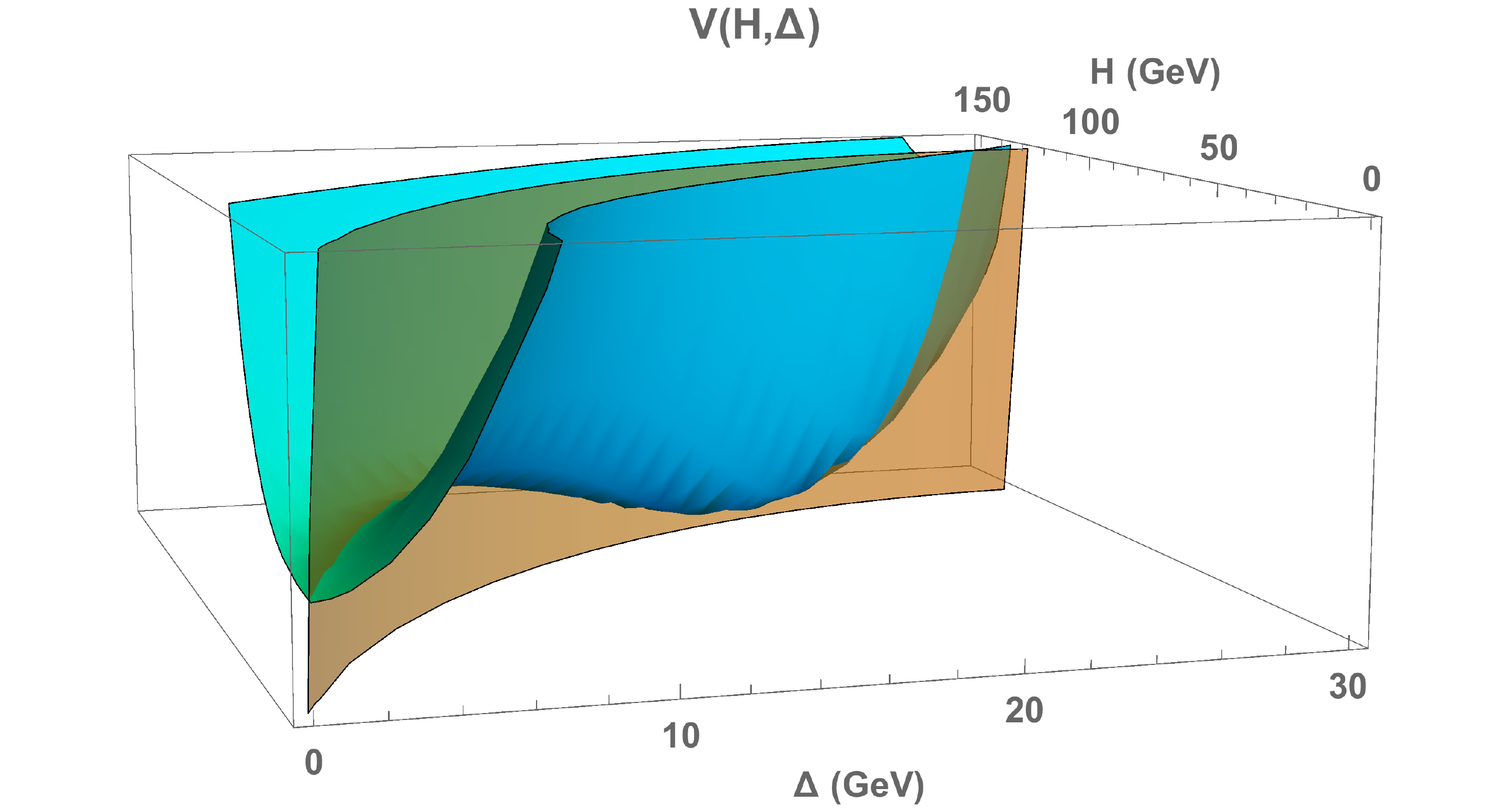}
\end{center}
\caption{~\emph{Two dimensional projection of the tree level potential in a point which exhibits a first order phase transition between the origin and the EW minimum (which for the considered point is located at $v_\Delta=20$ GeV and $v_H=116.35$ GeV). The orange plane that intersects the potential corresponds to the ellipsoidal direction that joins the origin and the EW minimum.}}
\label{fig:3dplot}
\end{figure}

In order to go from the two field configuration $(H_2^0, \phi^0)$ to one direction we will further reduce our field space by considering the smooth direction that joins the origin and the electroweak minimum passing through the saddle point~\footnote{As pointed out in~\cite{Coleman:1977py}, the tunneling path is the one where the barrier is minimized so any approximation will only overestimate the size of it.}, as can be seen in Fig.~\ref{fig:3dplot}. We have chosen this direction by considering an ellipse in the $(H_2^0, \phi^0)$ plane,
\be
H_2^0 \rightarrow f(\phi^0)= \left(1-a+ \sqrt{a^2+(a-1)^2-\left(\frac{\phi^0}{v_\phi}-a\right)^2}\right) v_2(T) 
\label{aprox2}
\ee
where the parameter $a$ is the eccentricity of the ellipse. By tuning $a$ we can get the right path and ensure that we connect smoothly the origin, the saddle point and the EW minimum at any temperature.  
 
The tunneling probability per unit time and unit volume from the false (symmetric) to the real (broken) minimum in a thermal bath is given by~\cite{Linde:1981zj},
\be
\frac{\Gamma}{\nu}\sim A(T)\,\mathrm{exp}\,[-B(T)],\quad B(T)\equiv \frac{S_3(T)}{T}
\ee
where the prefactor is $A(T)\simeq T^4$ and $S_3$ is the three-dimensional effective action. At very high temperature the bounce solution has $O(3)$ symmetry and the euclidean action is simplified to
\be
\label{s3}
S_3 = 4\pi \int_0^{\infty}r^2 dr \left[ \sum_k \frac{1}{2}\left(\frac{d\phi_k}{dr}\right)^2+V(\phi_k,T) \right] \, ,
\ee
where $r^2=\vec{x}^{\,2}$. Using (\ref{aprox1}) and (\ref{aprox2}) we can rewrite it as,
\be
S_3 = 4\pi \int_0^{\infty}r^2 dr \left[\frac{1}{2}F(\phi^0)\left(\frac{d\phi^0}{dr}\right)^2+V(\phi^0,T) \right]
\ee
where,
\be
F(\phi^0)= \left( 1+\frac{v_1(T)^2}{v_2(T)^2} \right)f'(\phi^0) +\left( 1+\frac{v_\psi(T)^2}{v_\phi(T)^2} +\frac{v_\chi(T)^2}{v_\phi(T)^2}\right)\, .
\ee

The bounce will be the solution to the euclidean equations of motion which yield the following equation
\be
F(\phi^0)\left[ \frac{d^2\phi^0}{dr^2}+\frac{2}{r}\frac{d\phi^0}{dr}\right] +\frac{1}{2} F'(\phi^0)\left( \frac{d\phi^0}{dr}\right)^2=V^\prime (\phi^0,T) \, ,
\ee
with the boundary conditions 
\be
\lim_{r\rightarrow \infty}\phi(r) = 0\quad \mathrm{and} \quad d\phi/dr|_{r=0}=0.
\ee
The nucleation temperature $T_n$ is defined as the temperature at which the probability for a bubble to be nucleated inside a horizon volume is of order one, in our case it turns out to happen when $S_3(T_n)/T_n\sim 135$. 

\begin{figure}[htb]
\begin{center}
\includegraphics[scale=.5]{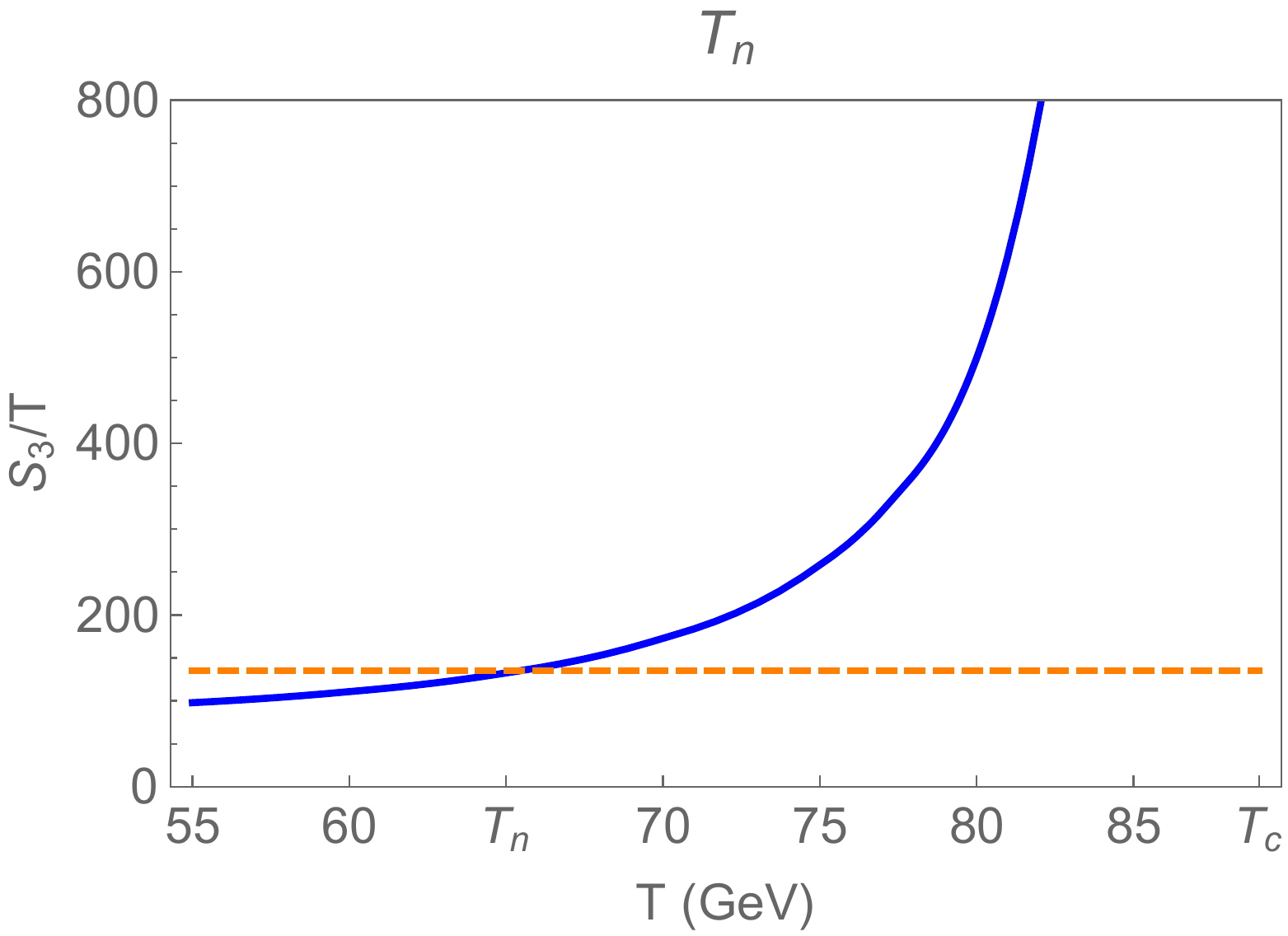}
\includegraphics[scale=.5]{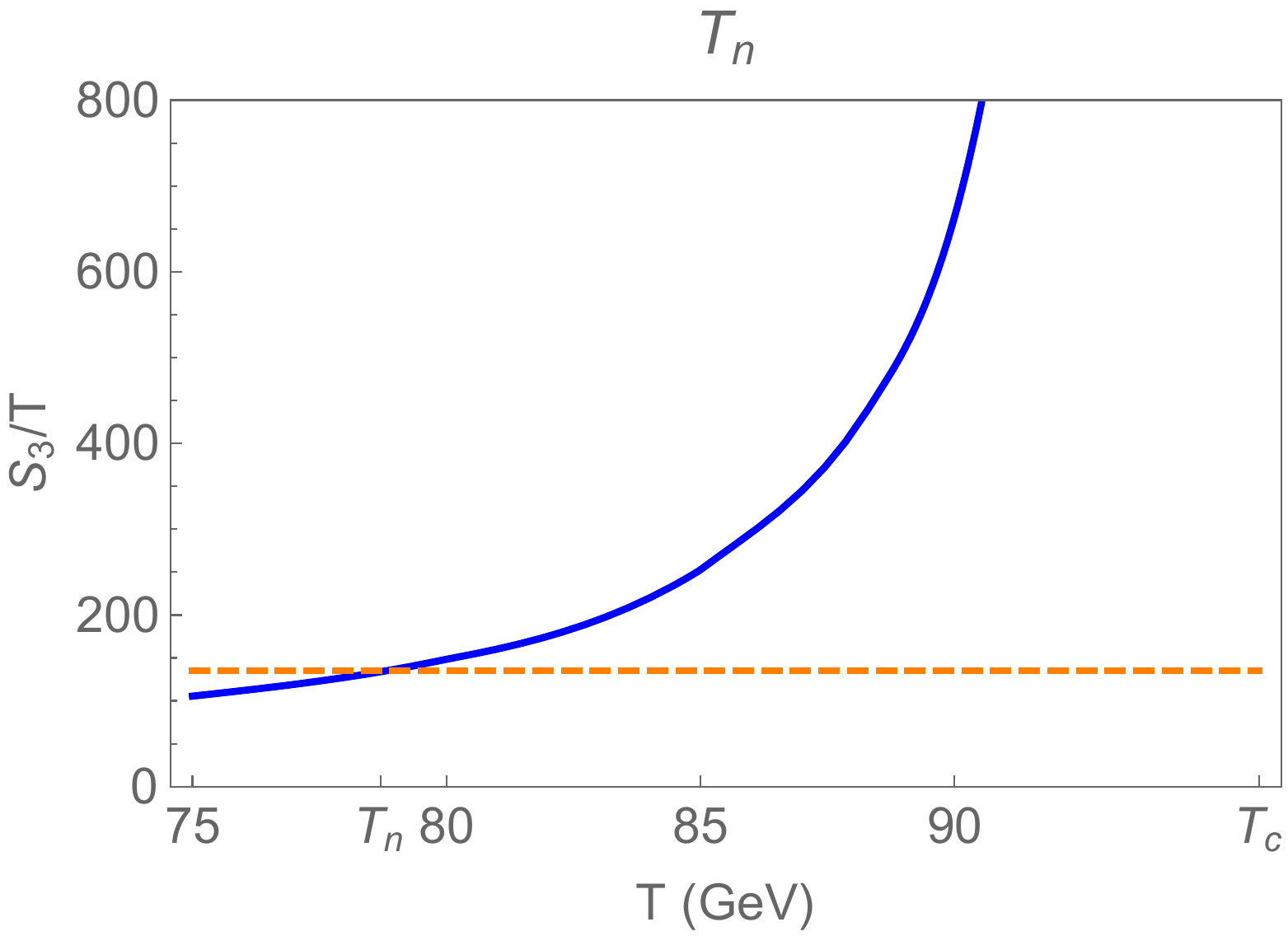}
\end{center}
\caption{~\emph{Left panel: Plot of the effective action over the temperature for $\mu=650$ GeV, $\mu_\Delta=475$ GeV, $v_\Delta=20$ GeV and $\tan{\beta}=1$. The order parameter at the critical temperature is $\phi(T_c)/T_c=1.82$ and $T_c=89$ GeV. The dashed line corresponds to $S_3(T)/T\sim 135$ and the crossing point with the thick blue line happens at the nucleation temperature $T_n=65$ GeV.  Right panel: The same for $\mu=650$ GeV, $\mu_\Delta=375$ GeV, $v_\Delta=20$ GeV and $\tan{\beta}=1.5$, where $\phi(T_c)/T_c=1.65$, $T_c=96$ GeV and $T_n=79$ GeV.}}
\label{fig:Tn}
\end{figure}

In Fig.~\ref{fig:Tn} we plot the effective action over the temperature for two points of the $(\mu,\mu_\Delta)$ plane. These plots show how the nucleation temperature depends on the strength of the phase transition. If the phase transition is not very strong then there is no large gap between the $T_n$ and $T_c$ (right plot). When the phase transition is very strong, a supercooling phenomenon happens and the nucleation temperature is quite smaller than $T_c$ (left plot in the figure). Of course if we move in the parameter space to points where $\phi(T_c)/T_c$ is even larger we will eventually find a situation where $S_3/T$ never reaches the correct value and bubble nucleation does not happen as the universe cools down. These points correspond to a thin band that is located at the bottom of the blue and red bands that we plot in Fig.~\ref{fig:firstorder}.

%%%%%%%%%%%%%%%%%%%%%%%%%%%%%%%%%%
%%%%%%%%%%%%%%%%%%%%%%%%%%%%%%%%%%
\section{Gravitational waves from the phase transition} \label{sec:pheno} 
\begin{figure}[htb]
\begin{center}
\includegraphics[scale=.5]{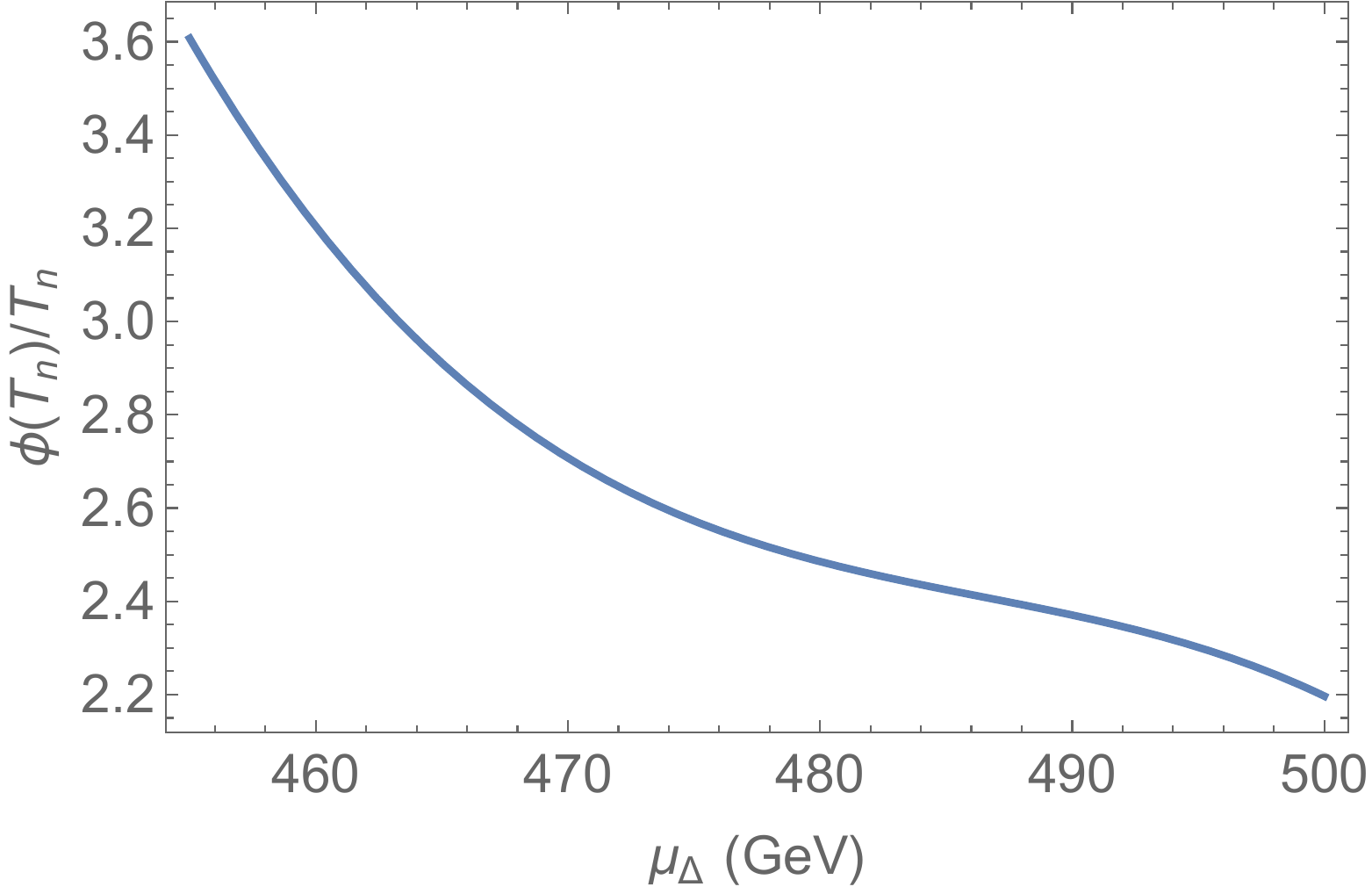}
\includegraphics[scale=.5]{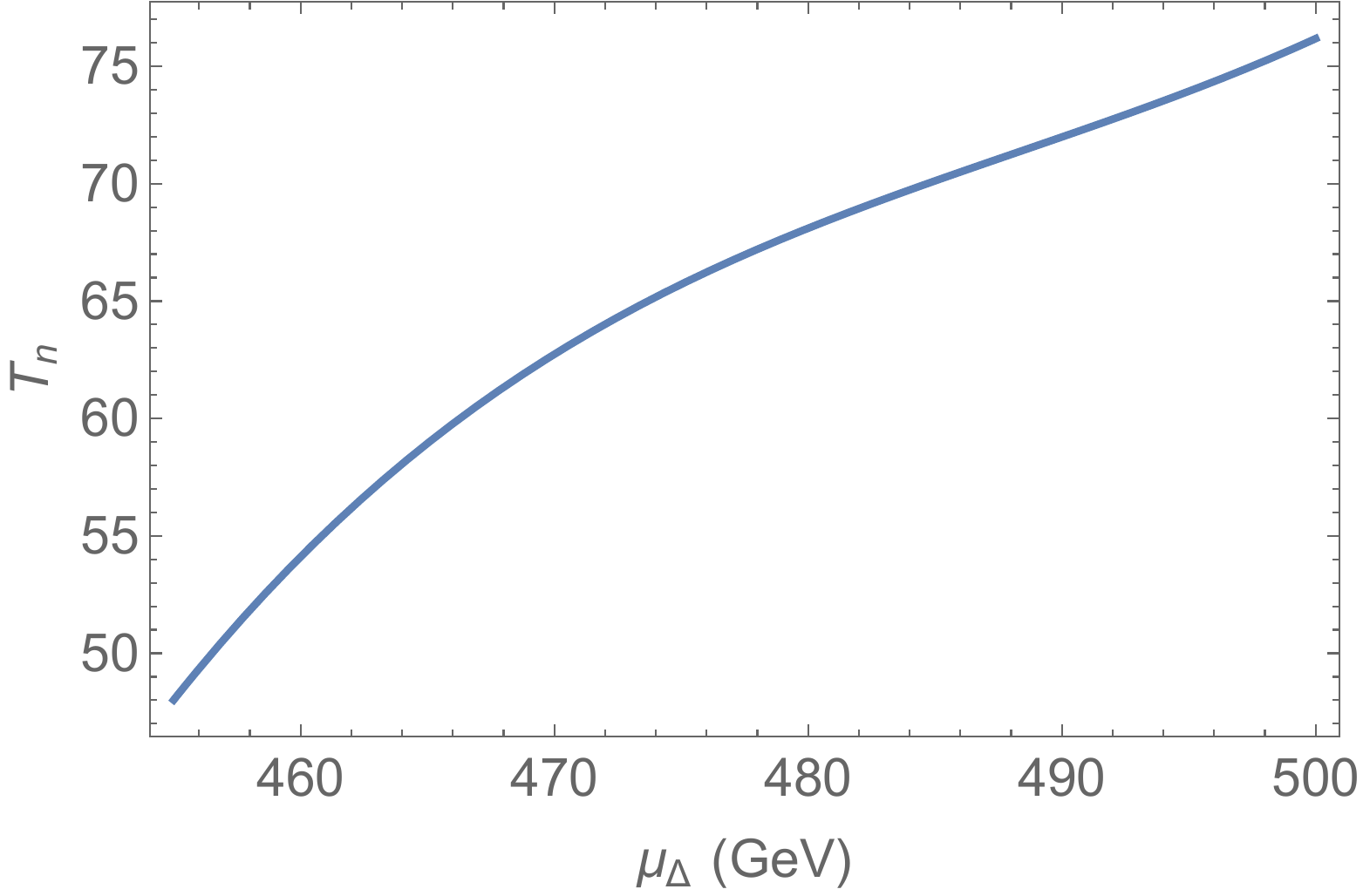}
\end{center}
\caption{~\emph{Left panel: Values of the order parameter at the nucleation temperature as a function of $\mu_\Delta$ for $v_\Delta=20$ GeV, $\mu=650$ GeV and $\tan{\beta}=1$. Right panel: Nucleation temperature $T_n$ for the same values of the parameters.}}
\label{fig:orderTn}
\end{figure}

It is known that a strong enough first order phase transition can generate sizable gravitational waves (GW). Since we are able to generate such a strong phase transition, due to the tree level nature of the barrier, we analyze in this section the possible spectrum of GWs. The spectrum can be characterized by only two parameters: the duration of the phase transition $1/\beta$, which is given by
\be
\frac{\beta}{H}=T\frac{d}{dT}\left(\frac{S_3}{T} \right)\ ,
\ee
and the latent heat
\be
\epsilon = \Delta V(T_n) -T_n \frac{d\, \Delta V(T)}{d T}\Big|_{T_n} \, ,
\ee
where
\be
\Delta V(T) = V(0,T)-V(\langle\phi(T)\rangle,T) \, .
\ee
The latent heat is usually normalized to the energy density of the radiation in the plasma, through the dimensionless parameter $\alpha$,
\be
\alpha = \frac{\epsilon}{\frac{\pi^2}{30}g_* T_n^4}
\ee
where $g_*$ is the effective number of degrees of freedom at the temperature $T_n$.
\begin{figure}[htb]
\begin{center}
\includegraphics[scale=.5]{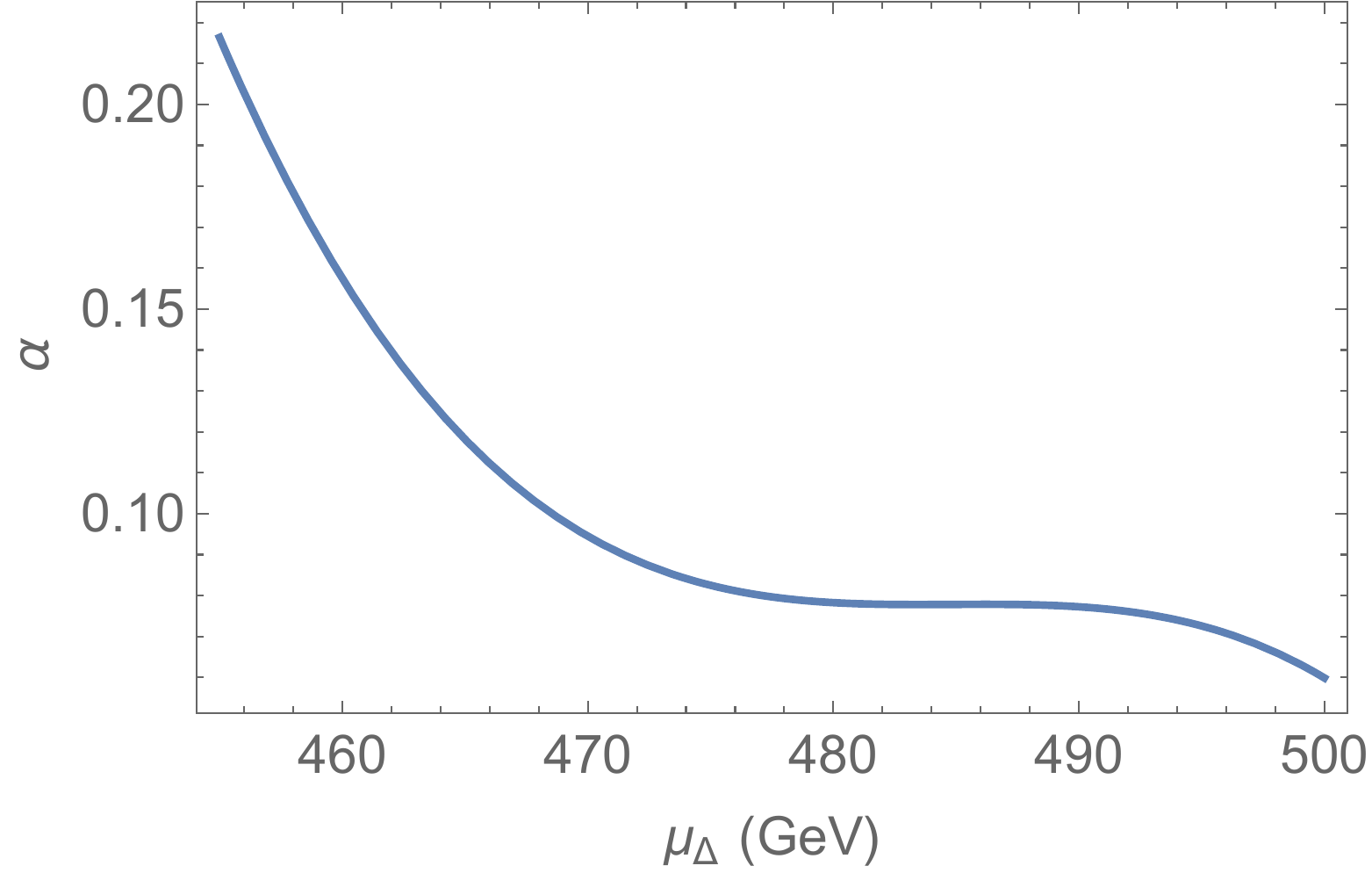}
\includegraphics[scale=.5]{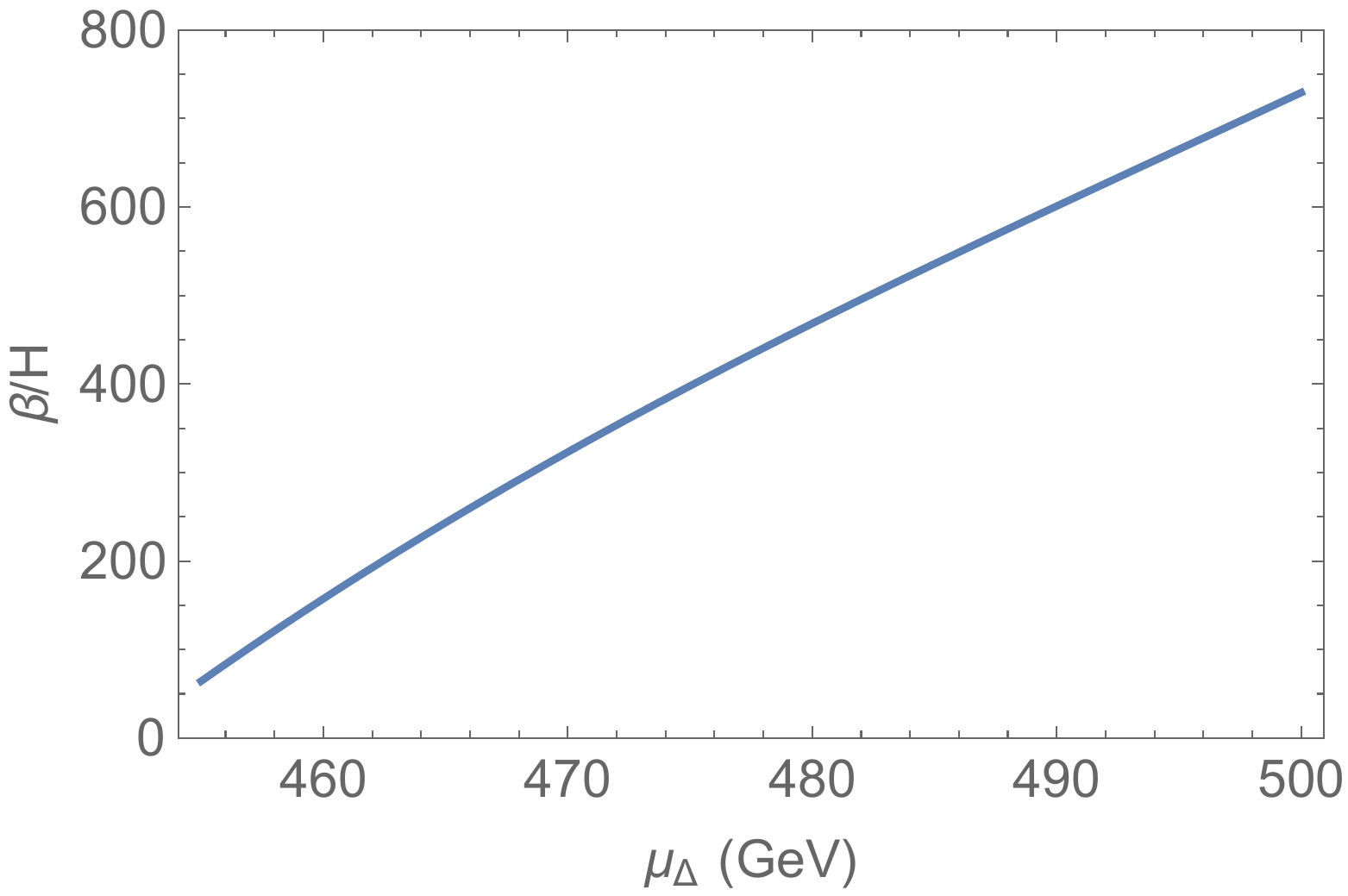}
\end{center}
\caption{~\emph{Left panel: Values of the $\alpha$ parameter for $v_\Delta=20$ GeV, $\mu=650$ GeV and $\tan{\beta}=1$. The number of effective degrees of freedom at the time of nucleation is $g_*=115.75$ . Right panel: Values of the $\beta/H$ parameter for the same values of the model parameters.}}
\label{fig:alphabeta}
\end{figure}
In Fig.~\ref{fig:alphabeta} we show results for the computation of the $\alpha$ (left panel) and $\beta/H$ (right panel) parameters along a vertical straight line of the band in Fig.~\ref{fig:firstorder} which corresponds to a fixed $\mu=650$ GeV value. 
In Fig.~\ref{fig:orderTn} we also show 
the values of the nucleation temperature (right panel) and the order parameter at that temperature (left panel). Note that for stronger values of the phase transition, $\alpha$ gets bigger and $\beta / H$ smaller. This means that the energy gap between the false and the true vacuum is big at the nucleation temperature and that the phase transition happens fast, which is precisely what one needs to get observable gravitational waves.

The above described parameters, which only depend on the finite temperature effective potential, are the only input coming from the particle physics model. Once we determine these two, we have to plug them into the cosmological picture. First we will treat the expanding bubbles, and the fluid they drag with, as if the bubbles where the only existing object. The collisions of these vacuum bubbles will then generate a GW spectrum~\cite{Huber:2008hg} (see Sec.~\ref{bubcol}). In the second part we will consider calculations that model the fluid in a more detailed manner, in this case the phase transition leads to the creation of sound waves which in turn will produce gravitational waves~\cite{Hindmarsh:2013xza} (see Sec.~\ref{soundwaves}).  

\subsection{Gravitational waves from bubble collisions}
\label{bubcol}
In the case we use the envelope approximation to model bubble collisions the peak frequency is~\cite{Huber:2008hg}
\be
\tilde{f}_{env}=16.5\,\mu \mathrm{Hz}\left( \frac{f}{\beta}\right) \left( \frac{\beta}{H}\right)\left( \frac{T_n}{100\,\mathrm{GeV}}\right)\left(\frac{g_*}{100}\right)^{1/6}\,
\ee
and the energy density
\be
h^2\tilde{\Omega}_{env}=1.84\times 10^{-6} \kappa^2\left( \frac{ v_b^3}{0.42 + v_b^2}\right)\left( \frac{H}{\beta}\right)^2\left( \frac{\alpha}{\alpha +1}\right)^2\left( \frac{100}{g_*}\right)^{1/3} \, .
\ee
The efficiency factor $\kappa$ is
\be
\kappa = \frac{1}{1+0.715\alpha}\left( 0.715\alpha +\frac{4}{27}\sqrt{\frac{3\alpha}{2}}\right)
\ee
the bubble wall velocity $v_b$ is
\be
v_b=\frac{\sqrt{1/3}+\sqrt{\alpha^2+2\alpha/3}}{1+\alpha}\, ,
\ee
and
\be
\frac{f}{\beta}=\frac{0.62}{1.8-0.1v_b+v_b^2}\, .
\ee
The spectrum then has the following shape
\be
\Omega_{env}(f)=\tilde{\Omega}_{env}\frac{3.8(f/\tilde{f}_{env})^{2.8}}{2.8+(f/\tilde{f}_{env})^{3.8}}\, .
\ee

\subsection{Gravitational waves from sound waves}
\label{soundwaves}
The peak amplitude of GW radiation from sound waves is given by~\cite{Hindmarsh:2013xza,Hindmarsh:2015qta}
\be
h^2\tilde{\Omega}_{sw}=2.65\cdot 10^{-6}\, v_b\, \kappa^2 \left(\frac{H}{\beta}\right)\left( \frac{\alpha}{\alpha +1}\right)^2 \left(\frac{g_*}{100}\right)^{-1/3} \, ,
\ee
which is larger than the result one gets from the envelope approximation by a factor $\beta/H$. The peak frequency is
\be
\tilde{f}_{sw}=19\, \mu\mathrm{Hz}\, \frac{1}{v_b}\left(\frac{\beta}{H}\right)\left(\frac{T_n}{100\, \mathrm{GeV}}\right)\left( \frac{g_*}{100}\right)^{1/6} 
\ee
and the fit to the numerical spectrum is given by
\be
\Omega_{sw}(f)=\tilde{\Omega}_{sw}\left( \frac{7}{4+3(f/\tilde{f}_{sw})^2}\right)^{7/2}(f/\tilde{f}_{sw})^3\, .
\ee

\subsection{Results for the spectrum of GWs}
\begin{figure}[htb]
\begin{center}
\includegraphics[scale=.7]{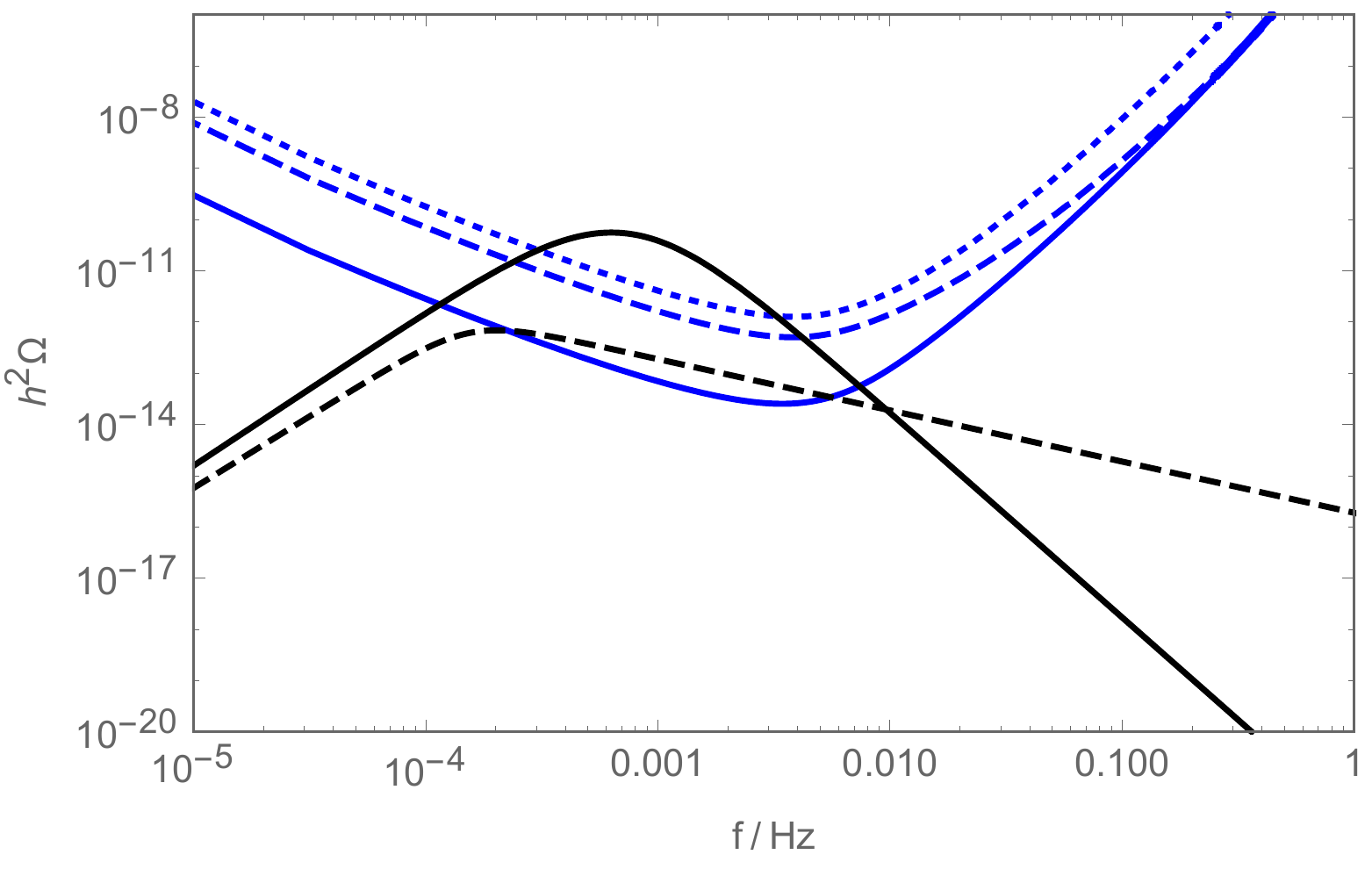}
\end{center}
\caption{~\emph{Spectrum of stochastic gravitational wave background coming from bubble collisions (dashed) and sound waves (solid) for a point where $\alpha\sim 0.22$ and $\beta/H\sim 57$, which corresponds to $\mu=650$ GeV and $\mu_\Delta=455$ GeV. The sensitivity curves of the eLISA designs are displayed in blue; design 1 (dotted), design 2 (dashed) and design 3 (solid).}}
\label{fig:spectrumGW}
\end{figure}

As we said in the previous section, when the phase transition is not radiatively generated, there will be points in the parameter space where the barrier is so large that no nucleation is possible. It is precisely close to these regions, but inside the region where the nucleation still happens, where the characteristics of the phase transition will be optimized for the detection of its GW spectrum. In particular the parameter $\beta/H$, will be minimized close to the region where $S_3/T$ never reaches the value $\sim 135$ and $\beta/H\sim 0$. As can be seen in Fig.~\ref{fig:alphabeta}, approaching this region we have found points where $\beta/H\sim 50$ and $\alpha\sim 0.22$. A spectrum coming from a point of these features is shown in Fig.~\ref{fig:spectrumGW} and may be probed by eLISA~\cite{Caprini:2015zlo,Huber:2015znp} and BBO~\cite{Corbin:2005ny,Harry:2006fi}. In the case of eLISA, the chances for detecting GWs improve with the design. Design 3, which features three $5$ Gm arms and $5$ years of data taking, is the one that could probe both GWs coming from bubble collisions, in the envelope approximation, and GWs coming from sound waves. We also see that GWs from sound waves could be detected by eLISA, even with design 1.

%%%%%%%%%%%%%%%%%%%%%%%%%%%%%%%%%%%%%%%%%
%%%%%%%%%%%%%%%%%%%%%%%%%%%%%%%%%%%%%%%%%
\section{Summary and Conclusions}
In this paper we have explored the nature of the EWPT in the SCTM. We have shown that, thanks to a tree level effect by which there is a barrier separating the minimum at the origin and the EWSB minimum, an important part of the parameter space of the model exhibits a phase transition whose order parameter is strong enough, both for the purpose of EWBG and for the detection of gravitational waves. We have decided to not focus on the regions where no barrier is generated at tree level (above the bands in Figs.~\ref{fig:firstorder} and~\ref{fig:lambda}), as analyzing the phase transition in this region would involve the consideration of higher order loop corrections in the thermal effective potential, which goes beyond the scope of the present paper.

In Sec.~\ref{sec:strength} we have discussed how the appearance of the barrier is directly linked to a non negligible contribution of the triplet sector to EWSB. Thanks to previous studies we can establish a relation between strong EWPT and collider searches. In fact the consequences for collider phenomenology of a scenario where EWSB is driven by doublets, but also features some triplet impurities, have already been studied in~\cite{Garcia-Pepin:2014yfa,Delgado:2015bwa}. In these papers a relation between a sizable $v_\Delta$ and light triplet like states was found, in agreement with the upper bounds derived in~\cite{Comelli:1996xg}. One therefore expects these new states to be there in the regions where a barrier is generated at tree level. As explained in the previous studies their detection is challenging due to their triplet like nature. However, modified Higgs coupling rates ($h\rightarrow\gamma\gamma$) or some signals such as $W^\pm W^\pm$ or $W^\pm Z$, which are specific of Higgs sectors with triplet representations, could act as smoking gun signals of the model and therefore probe the nature of the phase transition at high temperature.

We also have checked that nucleation does happen in most parts of the parameter space where the order parameter is larger than one. The potential of the model features a five-dimensional field space due to the introduction of three new triplet chiral superfields, on top of the two usual MSSM doublets. To simplify the calculation of the nucleation temperature we have minimized the euclidean action functional in the multi-field configuration space by using a smooth path going from the minimum at the origin to the EWSB minimum at finite temperature through the saddle point. Because of the character of our parameter space we are confident enough that the approximation works properly up to small corrections. In the last section it is shown how future interferometers such as eLISA could observe gravitational waves generated during the phase transition for some parts of the parameter space.

\section*{Acknowledgments} 
We thank Germano Nardini for useful discussions and informations about gravitational waves  and future interferometers. The work of M.G.-P. and M.Q.~is partly supported by MINECO under Grants CICYT-FEDER-FPA2011-25948 and
CICYT-FEDER-FPA2014-55613-P, by the Severo Ochoa Excellence Program of
MINECO under Grant SO-2012-0234 and by \textit{Secretaria d'Universitats i
Recerca del Departament d'Economia i Coneixement de la Generalitat de
Catalunya} under Grant 2014 SGR 1450.

%\clearpage

\bibliographystyle{JHEP}
\bibliography{refs}

\end{document}